\documentclass[journal]{IEEEtran}

\usepackage{cite}
\usepackage{mathtools}
\usepackage{amssymb}
\usepackage{amsmath}

\usepackage{graphicx}
\graphicspath{{converted_graphics/}}

\usepackage{subfig}

\DeclarePairedDelimiter\abs{\lvert}{\rvert}
\makeatletter
\let\oldabs\abs
\def\abs{\@ifstar{\oldabs}{\oldabs*}}

\usepackage{wrapfig}
\usepackage{parskip}
\usepackage[font={footnotesize}]{caption}
\usepackage{framed}

\def\footnoterule{\kern-3\p@
  \hrule \@width 2in \kern 2.6\p@}

\usepackage{amsfonts}

\def\footnoterule{\kern-3\p@
  \hrule \@width 2in \kern 2.6\p@}

\begin{document}

\title{Considerations on Quantum-Based\\
       Methods for Communication Security}

\author{{\large Jeffrey Uhlmann}\\
           Dept.\ of Electrical Engineering \& Computer Science\\
           University of Missouri-Columbia}

\markboth{NATIONAL ACADEMY OF SCIENCES, QUANTUM SENSING AND COMMUNICATIONS COLLOQUIUM, AUG.\ 23,~2018, WASHINGTON, DC.}{}

\maketitle
\thispagestyle{empty}

\begin{abstract}
In this paper we provide an intuitive-level discussion of 
the challenges and opportunities offered by quantum-based
methods for supporting secure communications, e.g., over
a network. The goal is to distill down to the most fundamental
issues and concepts in order to provide a clear foundation for 
assessing the potential value of quantum-based technologies 
relative to classical alternatives. It is hoped that this form of 
exposition can provide greater clarity of perspective than is
typically offered by mathematically-focused treatments
of the topic. It is also hoped that this clarity extends to
more general applications of quantum information science 
such as quantum computing and quantum sensing.
\let\thefootnote\relax\footnote{{\em Presentation to the National Academy of
Sciences, Quantum Sensing and Communications Colloquium, 
Washington, DC, August 23, 2018.}}
\addtocounter{footnote}{-1}\let\thefootnote\svthefootnote
\end{abstract}

\begin{IEEEkeywords}
Communication Security; Cryptography; One-Time Pad; 
Quantum Computing; Quantum Information; Quantum-Key Distribution;
Quantum Networks; Quantum Sensing; QKD.
\end{IEEEkeywords}

\section{Introduction}

Quantum-based technologies exploit physical phenomena that 
cannot be efficiently exhibited or simulated using technologies
that exploit purely classical physics.  
For example, a quantum sensor may use quantum
phenomena to probe a system to discern classical and/or quantum
properties of the system that cannot be directly measured by
classical sensing technologies. Quantum computing, by contrast,
generalizes the classical unit of information, the {\em bit}, in the
form of a quantum bit, or {\em qubit}, and exploits quantum
computational operators that cannot be efficiently simulated
using classical Boolean-based operators. 

Secure quantum-based communication protocols have emerged
as among the first practical technologies for which advantages 
over classical alternatives have been rigorously demonstrated.
As will be discussed, however, these advantages rely on a
set of assumptions about the capabilities of potential
adversaries (hackers) as well as those of the communicating parties.
Because the quantum advantage can be lost if these assumptions
are relaxed, the utility of quantum-based communication
must be assessed based on the assumed scenario in which it
will be applied.

In the next section we discuss scenarios in which 
classical cryptography can facilitate unconditionally secure
communications. We then discuss a more general
class of communication scenarios in which classical
methods cannot provide unconditional guarantees of
security but may offer practically sufficient ones. 
We then provide a high-level description of how
special properties of quantum systems can be
exploited to enlarge the range of scenarios for
which unconditional communication security can
be achieved. This provides context for realistically
examining how the tantalizing theoretical features
of quantum-based approaches to communication
security may translate to practical advantages over
classical alternatives.

\section{Secure Communications}
 
Suppose two parties, Alice and Bob, know they will have need
for unconditionally secure communications at various times
in the future. If they determine that they are unlikely to 
communicate more than a total of $n$ bits before the next
time they meet then they can create a sequence of 
random bits, referred to as a {\em one-time pad} (OTP), and 
each keep a copy for use to mask their messages. 

For example, a week later Alice can
contact Bob using whatever unsecure communication
medium she chooses, e.g., phone or email, and then
send her $k$-bit private message encrypted by performing
an exclusive-or\footnote{The exclusive-or function of two
bits $a$ and $b$ is $0$ if are the same and $1$ if they differ.}  
(XOR) of it with the first $k$ bits of the
OTP. Upon receipt of the encrypted k-bit message, Bob
will simply invert the mask by applying the same XOR
operation using the first $k$ bits of the OTP. 

Even if an eavesdropper, Eve, is able to monitor all 
communications between Alice and Bob, she will not be
able to access the private information (i.e., original 
plain-text messages) without a copy of their OTP.
Thus, the OTP protocol offers unconditional security
against eavesdropping, but its use is limited to
parties who have previously established a shared OTP.
The question is whether a secure protocol can be 
established between two parties who have never 
communicated before. 

\section{Public Key Encryption}

At first glance it appears that there is no 
way for Alice and Bob to communicate for
the first time in a way that is secure against
an eavesdropper who has access to every
bit of information they exchange. However,
a commonly-used analogy can quickly convey
how this might be done.

Suppose Alice wishes to mail a piece of paper
containing a secret message to Bob. To ensure
security during transport she places the paper
in a box and applies a lock before sending. When
Bob receives the box he of course can't open it
because of the lock, but he can apply his own lock
and send the box back to Alice. Upon receipt, 
Alice removes her lock and sends the box back
to Bob, who can now open it and read the 
message.

If it is assumed that the box and locks 
can't be compromised then this protocol
is secure even if Eve is able to gain
physical access to the box during transport.
An analogous protocol can
be applied to digital information if it
is possible for Alice and Bob to sequentially
encrypt a given message and then sequentially
decrypt it. To do so, however, Alice must be able 
to remove her encryption mask {\em after}
Bob has applied his. In other words, their
respective encryption operations must commute
and not be invertible by Eve. 

It turns out that no classical protocol
can satisfy the necessary properties for
unconditional security. However, a practical
equivalent of unconditional security 
can potentially be achieved in the sense 
that Eve may be able to invert the 
encryption -- but only if she expends
thousand years of computation time.
Under the assumption that security
of the message will be irrelevant at that
point in the distant future, the protocol
can be regarded as unconditionally secure 
{\em for all practical purposes}.
 
At present there are technically no
protocols that provably require such
large amounts of computational effort,
but some do if certain widely-believed
conjectures (relating to one-way functions)
are true. Assuming that these conjectures
are in fact true, classical 
public-key protocols would seem to
offer practically the same level of 
security as a one-time pad but without the
limitation of prior communication.

{\em On the other hand...} estimating
the expected amount of time necessary
to break a classical public-key protocol
is very difficult. Even if it is assumed that
the amount of work required by Eve grows 
exponentially with the length of a critical 
parameter, a particular value for that 
parameter must be chosen. For all existing
protocols the value of this parameter
introduces an overhead coefficient (both
in computational time and space) which
may not be exponential but may grow
such that the protocol
becomes impractical in most real-world
contexts.

Suppose the parameter is selected
based on a tradeoff between practical
constraints and a minimum acceptable
level of security, e.g., that it would take
Eve 500 years to break the encryption
using the fastest existing supercomputer.
What if Eve can apply 1000 supercomputers
and break it in six months? Or what if she
develops an optimized implementation of
the algorithm that is 1000 times faster? 
Breaking the code may still require time
that is exponential in the value of the 
parameter, but the real question is how
to estimate the range of parameter values 
that are at risk if Eve applies all available 
resources to crack a given message.

As an example, in 1977 it was estimated
that the time required to break a message
encrypted with the RSA public-key protocol 
using a particular parameter value would be 
on the order of many quadrillion years. However,
improved algorithms and computing resources
permitted messages of this kind to be broken
only four years later, and by 2005 it was
demonstrated that the same could be done
in only a day.

The difficulty of making predictions, especially
about the future\!~\cite{yogi}, raises significant 
doubts about the extent to which any particular 
classical public-key scheme truly provides a
desired level of security {\em for all
practical purposes}, and it is this nagging
concern that motivates interest in 
quantum-based protocols that 
offer true unconditional security, at
least in theory.

\section{Quantum Key Distribution (QKD)}

Quantum-based public-key protocols have been
developed that provide unconditional security
guaranteed by the laws of physics. In the case
of Quantum Key Distribution (QKD)\!~\cite{bb84}, 
its security is achieved
by exploiting properties that only hold for qubits.
The first is the {\em no-cloning theorem},
which says that the complete quantum
state of a qubit cannot be copied. 
The second is that a pair of qubits can be 
generated with {\em entangled} states such 
that the classical binary value measured for 
one by a particular measurement process using
parameter value $\Theta$ will be identical to what
is measured for the other using the same
parameter value, but not necessarily if
the second measurement is performed with a
different parameter value $\Theta'\neq\Theta$.

The no-cloning theorem is clearly non-classical
in the sense that a qubit stored in one variable 
can't be copied into a different variable the way 
the content of a classical binary variable can be
copied into another variable or to many other
variables. For example, if the
state of a given qubit is somehow placed into 
a different qubit then the state of
the original qubit will essentially be erased in 
the process\footnote{The theoretical physics
explaining why quantum states can't be cloned, 
and the details of how qubits are prepared and
manipulated, are not important in the present
context for the same reason that details of
how classical bits are implemented in 
semiconductor devices are not relevant
to discussions of algorithmic issues.}. 
In other words, the state of the
qubit should not be viewed as having been 
copied but rather {\em teleported} from the 
first qubit to the second qubit. If it is simply
measured, however, then its state collapses
to a classical bit and all subsequent measurements
will obtain the same result. 

Based on these properties, the following simple
quantum communication protocol\footnote{This
toy protocol is intended only to illuminate the
key concepts in a way 
that links to classical one-time pad (variations can be found
in\!~\cite{uhl15}). Much more complete
expositions of the general theory and practice of
quantum cryptography can be found 
in\!~\cite{assche,gisin,kollmitzer,sergienko}.} 
can be defined: 

\begin{enumerate}

\item Alice and Bob begin by agreeing on a set of 
        $k$ distinct measurement parameter values
        $\Theta=\{\Theta_1,\Theta_2,...,\Theta_k\}$.
        This is done openly without encryption, i.e.,
        Eve sees everything.
\item Alice and Bob each separately choose one of 
        the $k$ parameter values but do not communicate
        their choices, thus Eve has no knowledge of them.
\item Alice generates a pair of entangled qubits. She 
        measures one and sends the other to Bob. 
\item Bob reports his measured value. If Alice sees 
        that it is not the same as hers then she chooses
        a different parameter and repeats the process.
        She does this for each parameter value until only 
        one is found that always (for 
        a sufficiently large number of cases) yields the
        same measured value as Bob but does not
        give results expected for different $\Theta$ values.
\item At this point Alice and Bob have established a 
        shared parameter value that is unknown to Eve.
        The process can now be repeated to create
        a shared sequence of random bits that can be
        used like an ordinary one-time pad. 
        In fact, subsequent
        communications can be conducted securely
        using classical bits.
\end{enumerate}

The security of the above protocol derives from
the fact that Eve cannot clone $k$ copies of a 
given qubit to
measure with each $\Theta_k$, and simply
measuring transmitted qubits will prevent Alice 
and Bob from identifying a unique
shared measurement parameter. 
In other words, Eve may corrupt
the communication channel but cannot 
compromise its information.
At this point Alice and Bob can create 
a shared OTP (which they can verify
are identical by using a checksum or 
other indicator) and communicate
with a level of security beyond what is possible
for any classical public-key protocol.

\section{The Authentication Challenge}

For research purposes it is natural to 
introduce simplifying assumptions to
make a challenging problem more tractable.
The hope is that a solution to the simplified
problem will provide insights for solving 
the more complex variants that arise in
real-world applications. This was true of 
the lockbox example in which it was assumed
that Eve might obtain physical access to the
locked box but is not able to dismantle
and reassemble the box, or pick the lock,
to access the message inside. The secure
digital communication problem as posed in 
this paper also has such assumptions.

Up to now it has been assumed that 
Eve has enormous computational
resources at her disposal sufficient to
overcome the exponential computational
complexity demanded to break classical
protocols. Despite these resources, it
has also been assumed that she is only
able to {\em passively} monitor the channel
between Alice and Bob. This is necessary
because otherwise she could insert
herself and pretend to be Alice when
communicating with Bob and pretend
to be Bob when communicating with
Alice. This is referred to as a
Man-In-The-Middle (MITM) attack, 
which exploits what is known as the
{\em authentication} problem.

To appreciate why there can be no general
countermeasure to MITM attacks, consider
the case of Eve monitoring all of Alice's 
outgoing communications. At some point Eve
sees that Alice is trying to achieve first-time
communication with a guy named Bob. 
Eve can intercept the messages intended
for Bob and pretend to be Bob as the two
initiate a secure quantum-based protocol. 
Pretending to be Alice, Eve does the same
with Bob. Now all unconditionally secure
communications involve Eve as a hidden
go-between agent. 

In many respects it might seem easier
to actively tap into a physical channel 
(e.g., optical fiber or copper wire) than
to passively extract information from a
bundle of fibers or wires within an
encased conduit, but of course it's
possible to add physical countermeasures
to limit Eve's ability to penetrate that
conduit. On the other hand,
if that can be done then it might seem
possible to do something similar to thwart
passive monitoring. 

Ultimately no quantum public-key protocol 
can be unconditionally secure without
a solution to the authentication problem.
Many schemes have been developed in
this regard, but ultimately they all rely on
additional assumptions and/or restrictions or 
else involve mechanisms that potentially
could facilitate a comparable level of security
using purely classical protocols. 

As an example, suppose a company called
Amasoft Lexicon (AL) creates a service in
which customers can login and communicate
with other registered customers such that
AL serves as a trusted intermediary to manage 
all issues relating to authentication. This may
involve use of passwords, confirmation emails or
text messages to phones, etc., but ultimately 
it must rely on information that was privately 
established at some point between itself and 
each of its customers, e.g., Alice and Bob. 

Suppose each customer is required
to set up a strong password. Initially, how is 
that information exchanged securely with AL?
One option might be to require the customer
to physically visit a local provider so that 
the person's identity can be verified, and
a secure password can be established,
without having to go through an unsecure
channel. Okay, but how long must the 
password be? If it is to be repeatedly 
used then it would become increasingly
vulnerable as Eve monitors more and
more messages. 

To avoid repeated use of a short 
password, AL could give Alice a drive
containing 4TB of random bits for an
OTP that would be shared only by her
and AL. The
same would be done using a different
OTP when Bob registers. Now Alice
can initiate unconditionally secure 
communicates with AL, and AL can do
the same with Bob, and therefore
Alice and Bob can communicate with
unconditional security via AL. 

Regardless of whether communications
through AL involve a quantum component,
the security of the overall system depends
on the trusted security of AL -- and on the
security practices of its customers in
maintaining the integrity of their individual
OTPs. The situation can be viewed as
one of replacing one point of vulnerability
with a different one. For example, what
prevents Eve from seeking employment
at AL? Are there sufficient internal safeguards
to protect against nefarious actions of 
AL employees?

\section{The Complexity Challenge}

Complexity is a double-edged sword in
the context of communication security.
On the one hand it can be used to
increase the computational burden on 
Eve. On the 
other hand, it can introduce more points
of vulnerability for her to exploit as the
scale of the implementation (amount of
needed software and hardware) increases.

In the case of quantum-based protocols
there is need for highly complex infrastructure
to support the transmission of qubits and the
preservation of entangled states. The details
are beyond the scope of this paper, but it is
safe to say that as implementation details
become more concretely specified the 
number of identified practical vulnerabilities 
grows. 

An argument can be made that as long
as the theory is solid the engineering
challenges will eventually be surmounted.
This may be verified at some point in the
indefinite future, but it is worthwhile to consider 
the number of practical security challenges
that still exist in current web browsers,
operating systems, etc., despite the
recognized commercial and regulatory 
interests in addressing them.

The critical question is whether the
investment in quantum-based infrastructure
to support quantum-based secure 
communication protocols is analogous
to a homeowner wanting to improve
his security by installing a titanium
front door with sophisticated intruder
detection sensors but not making any
changes to windows and other doors.

The natural response to the titanium door analogy
is to agree that quantum-based technologies
represent only one part of the overall 
security solution and that of course there
are many other vulnerabilities which also
must be addressed. However, this raises
a new question: Is it possible that a complete
solution can be developed that doesn't 
require any quantum-based components?

It may turn out that it is only feasible to
guarantee {\em practically sufficient} levels
of security (as opposed to {\em unconditional})
and only for specialized infrastructure 
and protocols tailored to specific use-cases. 
If the scope of a given use-case is sufficiently
narrow (e.g., communications of financial 
information among a fixed number of
banks) then the prospects for confidently
establishing a desired level of security are
greatly improved. In other words, relative
simplicity tends to enhance trust in the
properties of a system because it is difficult 
to be fully confident about anything that is
too complicated to be fully understood.

\section{Discussion}

The foregoing considerations on the status
of quantum-based approaches for secure
communications have leaned strongly toward
a sober, devil's-advocate perspective\footnote{See
the appendices for more succinct expressions
of arguments considered in this paper.}. This
was intentional to firmly temper some of the
overly-enthusiastic depictions found in the
popular media. For example, the following is
from media coverage of an announcement in
May of 2017 about the launch of a quantum-based 
``unhackable'' fiber network in China:
\begin{quote}
{\em ``The particles cannot be destroyed or duplicated. 
Any eavesdropper will disrupt the entanglement and 
alert the authorities,'' a researcher at the Chinese 
Academy of Sciences is quoted as saying.}
\end{quote}
Hopefully our discussion thus far clarifies the
extent to which there is a factual basis for this
quote and how the implicit conclusion (i.e., that 
the network is ``unhackable'') goes somewhat
beyond that basis. One conclusion that cannot 
be doubted is that remarkable progress has
been made toward implementing practical systems
based on theoretically-proposed quantum 
techniques. Another equally-important
conclusion that can be 
drawn is that China is presently leading this 
progress. 

In many respects the situation is
similar to the early days of radar when it was
touted as a sensing modality that could not be
evaded by any aircraft or missile because it had
the means ``to see through clouds and darkness.'' 
While this claimed capability was not inaccurate,
that power motivated the development of increasingly
sophisticated countermeasures to mask the 
visibility of aircraft to enemy radar, thus 
motivating the development of increasingly more 
sophisticated technologies to counter those 
countermeasures. The lesson from this is that
every powerful technology will demand 
continuing research and development to
meet new challenges and to support new
applications.

It is likely that the real value of future
quantum fiber networks will not be communication
security but rather to support the needs of 
distributed quantum sensing applications.
More specifically, quantum information from
quantum-based sensors and related technologies 
can only be transmitted via special channels 
that are implemented to preserve entangled
quantum states. The future is quantum,
so the development of infrastructure to
manage and transmit quantum information
has to be among the highest of priorities.

\section{Conclusion}

In retrospect it seems almost ludicrous to suggest
that any technology could ever offer something as 
unequivocally absolute as ``unconditional guaranteed
security,'' but that doesn't mean quantum-based
technologies don't represent the future state-of-the-art 
for maximizing network communication security. 
More importantly, surmounting the theoretical and 
practical challenges required to realize this state-of-the-art 
will have much more profound implications than 
simply supporting the privacy concerns of Alice and
Bob.
\newpage
\begin{appendices}
\section{Devil's Advocate Arguments}
\begin{em}
\begin{itemize}
\item ``The theoretical guarantees provided by QKD are only satisfied under certain assumptions. It may be that those assumptions can't be satisfied in any practical implementation and thus QKD provides no theoretical advantages over classical alternatives.''

\item ``If it's possible to implement the highly-complex infrastructure needed to support QKD, and to provide physical security against MITM attacks, then it should also be possible to implement physical security against passive monitoring. If that can be achieved then there is no need for QKD.''

\item ``The complexity associated with QKD may make it less secure than simpler classical alternatives. Just consider the number of security challenges that still exist in current web browsers, operating systems, etc., despite the recognized commercial and regulatory interests in addressing them.''

\item ``Progress on the development of classical protocols (e.g., based on elliptic curve cryptography) may very well lead to rigorous guarantees about the asymmetric computational burden imposed on Eve. If so, this would provide essentially unconditional security for all practical purposes.''

\item ``The need for provable unconditional security may be limited to only a few relatively narrow contexts in which classical alternatives are sufficient. For example, communications of financial information among a fixed number of banks could potentially be supported using classical one-time pads that are jointly established at regular intervals.''

\item ``QKD assumptions on what the physical infrastructure is required to support, and on what Eve is and is not able to do, seem to evolve over time purely to conform to the limits of what the theoretical approach can accommodate. This raises further doubts about QKD's true scope of practical applicability.'' 

\item ``Implementing quantum infrastructure to support QKD is analogous to a homeowner wanting to improve security by installing a titanium front door but not making any changes to windows and other doors. In the case of Alice and Bob, for example, it's probably much easier for Eve to place malware on their computers, or place sensors at their homes, than to identify and compromise a network link somewhere between them.''
\end{itemize}
\end{em}

\section{Replies to the Devil's advocate:}
\begin{em}
\begin{itemize}
\item ``If demands are set too high at the outset then no progress can ever be made to improve the status quo.''

\item ``Even if it is true that most security-critical applications will demand specially-tailored solutions, the availability of quantum-based tools will offer greater flexibility in producing those solutions.''

\item ``People can assume responsibility for their local security but have no choice
but to trust the security of infrastructure outside their control.''

\item ``A network that supports quantum information is unquestionably more powerful than one that does not. It is impossible to foresee the many ways this power will be exploited down the road, but it is hard to imagine that enhanced security will not be included.''
\end{itemize}
\end{em}
\end{appendices}

\end{document}